\documentclass[useAMS,usenatbib]{mn2e}
\usepackage{times}
\usepackage{epsfig}
\usepackage{graphicx}
\usepackage{latexsym}
\usepackage{pifont}
\input epsf

\newcommand{\xmmn}{{\it XMM-Newton~\/}}
\newcommand{\asca}{{\it ASCA~\/}}
\newcommand{\chandra}{{\it Chandra~\/}}

\newcommand{\rosat}{{\it ROSAT~\/}}
\newcommand{\einstein}{{\it Einstein~\/}}

\def\ergcms{{\rm ~erg~cm^{-2}~s^{-1}}}

\def\ergsec{{\rm ~erg~s^{-1}}}
\def\cms{{\rm ~cm^{-2}}}

\def\H0{{\rm ~km~s^{-1}~Mpc^{-1}}}

\def\etal{et al.~\/}

\def\la{\mathrel{\hbox{\rlap{\hbox{\lower4pt\hbox{$\sim$}}}{\raise2pt\hbox{$<$}}}}}
\def\ga{\mathrel{\hbox{\rlap{\hbox{\lower4pt\hbox{$\sim$}}}{\raise2pt\hbox{$>$}}}}}
\def\d25{D$_{25}$}

\def\hii{H{\small II}$~$}

\def\deg{\hbox{$^\circ$~\/}}
\def\arcm{\hbox{$^\prime$~\/}}
\def\arcs{\hbox{$^{\prime\prime}$~\/}}

\def\eps@scaling{1.0}%
\newcommand\plottwo[2]{%
  \centering
  \leavevmode
  \columnwidth=.45\columnwidth
  \includegraphics[width={\eps@scaling\columnwidth}]{#1}%
  \hfil
  \includegraphics[width={\eps@scaling\columnwidth}]{#2}%
}%

\title[An \xmmn view of M101 - III] {An \xmmn view of M101 - III. Diffuse X-ray emission}
\author[R.S Warwick et al.]
	{R.S.\ Warwick$^{1}$, L.P.\ Jenkins$^{1,2}$, A.M.\ Read$^{1}$, 
T.P.\ Roberts$^{1,3}$, R.A.\ Owen$^{1}$ \\  
$^{1}$~X-ray \& Observational Astronomy Group, Dept. of Physics \& Astronomy, 
University of Leicester, University Road, Leicester LE1 7RH, U.K.\\
$^{2}$~Laboratory for X-ray Astrophysics, NASA Goddard Space Flight Center, 
Greenbelt, MD 20771, USA \\
$^{3}$~Department of Physics, Durham University, South Road, Durham DH1 3LE, UK \\}

\date{Accepted ......................; Received .....................; 
in original form .....................}

\pagerange{\pageref{firstpage}--\pageref{lastpage}}

\pubyear{2005}

\begin{document}

\maketitle

\label{firstpage}

\begin{abstract}


We present a study of the X-ray properties of the nearby face-on Scd spiral galaxy
M101 based on recent \xmmn observations.  In this third and final paper in the 
present series, we focus on the spatial and spectral properties of the residual
emission, after excluding bright X-ray sources with $L_X > 10^{37} \ergsec$.
Within a central region of radius 10\arcm (21 kpc), the X-ray emission broadly 
traces the pattern of the spiral arms, establishing a strong link with recent star 
formation, but it also exhibits a radial scale length of $\approx 2.6\arcm$ (5.4 kpc) 
consistent with optical data. We estimate the
soft X-ray luminosity within the central $5\arcm$ (10.5 kpc) region to
be $L_X \approx 2.1 \times 10^{39} \ergsec$ (0.5--2 keV), the bulk of which 
appears to originate as diffuse emission. We find a two-temperature thermal model best fits
the spectral data with derived temperatures of 0.20$^{+0.01}_{-0.01}$ keV and 
0.68$^{+0.06}_{-0.04}$ keV which are very typical of the diffuse components seen
in other normal and starburst galaxies. More detailed investigation
of the X-ray morphology reveals a strong correlation with images recorded
in the far-UV through to V band, with the best match being with the U band.
We interpret these results in terms of a clumpy thin-disk
component which traces the spiral arms of M101 plus an extended 
lower-halo component with large filling factor.

\end{abstract}


\section{Introduction}
\label{sec:intro}

The  hot ($T\sim10^6$\,K), X-ray  emitting plasma  observed in  the disk  of our
Galaxy is an  important factor when considering the  energy balance and enrichment
of the interstellar  medium (ISM). Energised by the winds of young massive stars and 
supernovae, its properties relate to the supernova rate, the evolution  of 
supernova remnants (SNRs) and the impact of young star clusters on their environment. 
In normal spiral galaxies like the Milky Way, hot gas produced by  multiple  
supernova  explosions may escape from the disk via galactic   fountains and/or chimneys 
(\citealt{shapiro76};  \citealt{bregman80}; \citealt{norman89}) 
resulting  in a dynamic corona that  is bound to the galaxy. This gas  then cools, 
condenses and falls back to  the disk, possibly as high-velocity clouds  
\citep{bregman80}. In contrast, the extremely high  star-formation rates in 
starburst  galaxies give rise to hotter ($T\ga10^7$\,K)  large-scale  metal-rich  
outflows  from the  starburst regions in the form of ``superwinds''
({\it e.g.,}  \citealt*{heckman90}). Such superwinds are energetic enough to fully escape 
the influence of the galaxy and are potentially important for the enrichment 
and heating of the intergalactic medium ({\it e.g.,}  \citealt*{martin02}).

Unfortunately,  the  opacity  of  the  galactic plane  to  soft  X-rays  greatly
complicates the study  of the extent, physical properties  and filling factor of
the  X-ray  emitting  plasma in our own galaxy.  However, observations  of  nearby  
spiral  galaxies  with
configurations  ranging from  face-on  to  edge-on can  circumvent  many of  the
problems inherent  in studies of the  Milky Way. The \einstein observatory provided 
the first detection of ultra-hot out-flows from the disks  of edge-on starburst
galaxies (\citealt{watson84}), although  only upper limits were obtained on
the  diffuse emission produced in the disks of normal spiral  galaxies 
(\citealt{bregman82};
\citealt{mccammon84}). The improved throughput  and spatial resolution
of  subsequent \rosat  PSPC/HRI observations  revealed unambiguous  evidence for
truly    diffuse     emission    in    spiral     galaxies    (\citealt{cui96};
\citealt{read97}). More recently, diffuse  structures have been observed with the
\xmmn    and   \chandra    observatories    in   samples    of   both    edge-on
(\citealt{pietsch01}; \citealt*{stevens03}; \citealt{strickland04a};
\citealt{strickland04b})   and  face-on   spiral   galaxies  (\citealt{kuntz03};
\citealt{tyler04}).  In  the  face-on  systems,  diffuse X-rays  were  found  to
correlate  with the  nuclear regions,  trace  recent star  formation within  the
spiral  arms and correlate  spatially with  H$\alpha$ and  mid-infrared emission
\citep{tyler04}.   In  the   survey  of   edge-on  systems   by   Strickland  et
al.  (2004a,b), extra-planar  diffuse  emission was  detected  in all  starburst
galaxies  and one  normal spiral  galaxy, the  extent of  which  correlated with
estimates of the rate of star formation within the disks.

\begin{table*}
\caption{Details of the three \xmmn observations of M101.}
 \centering
  \begin{tabular}{lcccccccccc}
\hline
\#    & Observation ID & Date         & & Filter   & & \multicolumn{2}{c}{Pointing co-ordinates}  & & \multicolumn{2}{c}{Useful exposure (ks)} \\
      &                & (yyyy-mm-dd) & &          & & RA (J2000)       & Dec (J2000)             & & pn            & MOS                      \\ 
\hline
1     & 0104260101     & 2002-06-04   & & Medium   & & $14^h03^m10.0^s$ & $+54\deg20\arcm24\arcs$ & & 25.7          & 36.7                     \\
2     & 0164560701     & 2004-07-23   & & Medium   & & $14^h03^m32.3^s$ & $+54\deg21\arcm03\arcs$ & & 26.2          & 28.9                     \\
3     & 0212480201     & 2005-01-08   & & Thin     & & $14^h03^m32.7^s$ & $+54\deg21\arcm02\arcs$ & & 27.7          & 25.8                     \\
\hline
\end{tabular}
\label{table:obs}
\end{table*}

Clearly, if  the purpose is to image  the X-ray emission over  a wide bandwidth,
one should preferentially select galaxies in
directions where the  line-of-sight Galactic hydrogen column density  is low. In
this   context,  the   Scd   supergiant  spiral   M101   (NGC~5457),  for   which
$N_H = 1.1\times10^{20}  \cms$ \citep{dickey90},  is an  ideal candidate  for a
face-on study.  A diffuse  X-ray component  was first hinted  at in  the central
regions   of   M101   with   \einstein  at   energies  in the range   0.2--1.5\,keV
(\citealt{mccammon84}; \citealt{trinchieri90}). Subsequent \rosat  PSPC
observations \citep{snowden95} revealed extensive  diffuse X-ray emission in the
0.1--1\,keV band, corresponding to thermal  emission   in  the  temperature   
range  1--3$\times10^6$\,K.  The
dominant soft component was  spatially  peaked towards  the
centre of  the galaxy  but could also  be traced out  to  a  radius  of  at least  
7\arcm (15\,kpc at  a distance of  7.2\,Mpc). It was  argued that 
this soft emission could not originate solely in the disk as this would require
a filling factor greater than unity.  After correcting for absorption intrinsic 
to M101, the total diffuse luminosity was estimated to be at least 
$10^{40} \ergsec$.

In  the \rosat  HRI  study of  \citet{wangetal99}, low-surface-brightness  X-ray
emission was  detected both in  the central region of M101  and in the vicinity  
of bright \hii regions in the  spiral arms of the galaxy. The improved 
spatial  resolution of the HRI
($\sim5$\arcs) allowed the detection and subtraction of point sources down
to  a  luminosity  of   $4\times10^{37}  \ergsec$,  after  which  the  total
0.5--2\,keV luminosity of the residual  emission within a radius of 12\arcm
was  estimated to  be $9\times10^{39}  \ergsec$. These  authors  agreed with
\citet{snowden95}  that faint discrete  X-ray sources  were likely to account
for only a fraction of the observed flux, implying that the bulk of the soft 
emission is truly diffuse in nature. The X-ray spectra
of the emission  in the central region  from both the \rosat PSPC  and \asca GIS
were fitted simultaneously, yielding a soft thermal component with a temperature
of $kT\sim0.2$\,keV, plus a hard power-law continuum most plausibly explained
as the integrated emission of the population of hard X-ray binary sources (XRBs)
in M101.

Further progress was  made in the \chandra study  of \citet{kuntz03}, using data
from the  8\arcm  square ACIS-S3 chip  from the  first $\sim$100\,ks
\chandra  observation  of  M101.  The  superb spatial  resolution  of  \chandra
allowed the removal of point sources down to luminosities of $\sim10^{36}
\ergsec$, leaving  residual   emission of luminosity
2.3$\times10^{39} \ergsec$ in the 0.45--1.0\,keV band, 16\% of which
was estimated to come from unresolved XRBs and dwarf stars. The bulk of the soft
emission was found  to trace the spiral arms and  was also spatially correlated
with H$\alpha$ and  far ultra-violet (UV) emission, physically  linking the soft
X-ray emission to  regions of on-going star formation. The  X-ray spectrum of the
diffuse  emission was well-described  by a  two-temperature thermal  plasma with
$kT$=0.2/0.75\,keV,  with a large  covering factor  implying that  a significant
fraction of the softest component originates in the halo of M101.

This is the third in a series of papers reporting the results of \xmmn observations  
of M101. In \citet{jenkins04a}~(hereafter Paper~I),  the spectral  and timing
properties   of  the   brightest  X-ray  sources   were  investigated, 
whereas in \citet{jenkins05b}~(hereafter  Paper~II) we reported on  the 
full catalogues of X-ray source detected in M101 by {\it XMM-Newton}.  
In the current 
paper we focus  on  the  morphology  and  spectral properties  of  the  
``diffuse''  X-ray component in this galaxy.
The structure of this paper is as follows.
First we describe the \xmmn observations and the methods used to 
construct images and spectra of the residual emission after the exclusion of bright
discrete sources (section \ref{sec:obs}). Next we investigate the likely 
composition, spectral properties and  spatial morphology of this residual
emission (section \ref{sec:res}). We then go on to discuss the implications of our 
results (section \ref{sec:disc}) and, finally, provide a brief summary of our conclusions 
(section \ref{sec:conc}). Throughout this paper we assume a distance to 
M101 of 7.2 Mpc (\citealt*{stetson98}), implying that an angular scale of 1\arcm 
corresponds to a linear extent of 2.09 kpc in M101.

\section{Observations \& Data Reduction}
\label{sec:obs}

Three  \xmmn  EPIC observations have been made of M101 as summarised in 
Table~\ref{table:obs}.  Our earlier analysis of the point source population 
in M101, presented in Papers~I and II of this series, utilised only the first
observation performed in 2002. However, for this paper, we have  
taken advantage of two additional observations carried out by
\xmmn in 2004/5 as a Target of Opportunity (TOO) programme to  follow  up  the  
outburst   of  the  ultraluminous  supersoft  X-ray  source J140332.3+542103,    
(see \citealt*{kong04}; \citealt*{mukai05}). The first \xmmn observation was 
targeted at the nucleus of the galaxy with the result that the 
30\arcm diameter field-of-view  of the EPIC 
cameras encompassed the full 28.8\arcm (\d25) extent of M101
(\citealt{devaucouleurs91}). In  the second and third observations the source 
J140332.3+542103 was placed  on axis, resulting in a $3.3'$  eastward offset 
of the field-of-view with respect to the first observation.

All datasets were screened for periods of high background by
accumulating full-field 10--15\,keV light curves. MOS data were 
excluded during periods when the 10-15 keV count rate exceeded 
0.3 $\rm ct~s^{-1}$, and the pn data were similarly screened 
when the pn count rate exceeded 3 $\rm ct~s^{-1}$ (except in 
observation~1, which was more heavily contaminated by  
flaring, where the pn threshold was set at 0.9 $\rm ct~s^{-1}$).  
For the subsequent image and spectral analysis, single and double  pixel 
events were selected for the pn (pattern 0--4), whereas single to quadruple 
events (pattern 0--12) were utilised in the case of the MOS data.

\subsection{Image processing and point source subtraction}

To study the morphology of the diffuse component, pn and MOS images and associated
exposure maps were created for each observation in  three energy bands: 
soft (0.3--1\,keV),  medium  (1--2\,keV)  and   hard  (2--6\,keV).
In each case the pixel size was set at $4.35'' \times 4.35''$.
The pn, MOS 1 and MOS 2 images in each of the three bands were first flat-fielded
by subtracting a constant (non-vignetted) background particle rate (estimated 
from the count rates recorded in the corners of the CCD detectors
not exposed to the sky) and then dividing by the appropriate exposure
map.  In the same process bad pixels, hot columns and spurious data along chip
gaps were excised.  The resulting data from the three observations were then 
mosaiced into flat-fielded pn and MOS (MOS 1 plus MOS 2) images for each band. 
Since the medium and hard band images showed no evidence for diffuse emission,
here we focus primarily on the images obtained in the soft (0.3-1 keV) band.
Figure~\ref{fig:im_big}(a) shows an adaptively-smoothed soft-band image, 
obtained from the combined pn and MOS datasets, encompassing the central 
20\arcm diameter region centred on the optical nucleus
of M101 (RA $\rm 14^h~3^m~12.55^s$, Dec $+54\deg~20\arcm~56.5\arcs$) .  The amplitude
scaling is logarithmic and adjusted so as to highlight the extended low surface 
brightness emission in M101.

\begin{figure*}
\centering
\caption{ {\it (a)} \xmmn soft-band (0.3-1 keV) image of the central $20'$ of M101. This 
flat-fielded image is constructed using pn and MOS data from the three \xmmn 
observations. The amplitude scaling is logarithmic with a dynamic range (maximum 
to minimum in the colour table) of 50. The position of the optical nucleus of 
M101  is marked with a cross. {\it (b)}  A ``model'' image constructed from a set of
count-rate scaled PSF sub-images.  A source mask, defined by the black contours,
was produced by applying a cut at a surface brightness threshold of 0.01 pn+MOS $\rm 
ct~ks^{-1}~pixel^{-1}$.  {\it (c)} The same X-ray image as shown in panel (a)
with the regions confused by bright sources masked out. The amplitude scaling is 
logarithmic with a dynamic range of 10. {\it (d)} The {\it GALEX} near-UV 
(NUV:2310 \AA) band image of M101 on the same spatial scale 
as the X-ray data. The  scaling is logarithmic with a dynamic range of 10.
In all four panels the two large circles have 
radii of $5'$ and $10'$.  {\tt Figures submitted to arXiv as jpgs.}
}
\label{fig:im_big}
\end{figure*}

To facilitate the investigation of the contribution of bright sources to the 
total X-ray emission in M101,  a master sourcelist, applicable to the three
observations, was created. To this end, each of the new observations was analysed in turn 
using the source detection procedure described in Paper II (applied to
the soft, medium and hard band data). Then, by comparing the resulting source lists, 
the 108 sources detected in observation 1 (Paper II) were supplemented by 21 new sources 
from observation 2 and an additional 9 new sources from observation 3, to 
give a combined source catalogue of 138 sources (see the Appendix for details of the 
new sources). 

The next step involved the construction of a spatial mask for the suppression of the
bright X-ray sources in the central field of M101. 
Using the pn+MOS flat-fielded soft-band image, a soft-band source count 
rate, net of the local background, was determined at each source position; in practice, 
we extracted the count rate within a cell of $16''$ radius centred on the source 
position, subtracted the local background, and lastly applied a correction for the extent 
of the point spread function (PSF) beyond the source cell.  We then selected the sources 
within 11\arcm radius of the nucleus of M101 with count rates above 
1.5 pn+MOS $\rm ct~ks^{-1}$ as a ``bright 
source sample'' to be excised from the the soft-band image. Here our approach was to 
construct a model image in which each source in our bright source list was 
represented by a PSF sub-image 
centred on the source position and an intensity scaling commensurate 
with the measured source count rate.  Then by applying a cut at a surface brightness 
threshold of 0.01 pn+MOS $\rm ct~ks^{-1}~pixel^{-1}$, we were able to construct 
a bright source mask - see  Fig.~\ref{fig:im_big}(b). A total of 65 bright soft-band
sources were removed within a radius of 11\arcm of the nucleus of M101 (25 sources
within 5\arcm).

Figure \ref{fig:im_big}(c) shows the pn+MOS soft-band image of M101 after applying the 
bright source mask
with the colour table adjusted to highlight the distribution
of the residual emission\footnote{Hereafter in this paper we use the word ``residual'' 
to describe the X-ray emission of M101 after masking out the bright 
discrete X-ray sources. This residual component will include the truly
diffuse emission, the integrated emission of discrete sources with count rates
below our source-exclusion threshold and also some low-level contamination from 
bright sources due to the spillover beyond the masked regions of their extended 
PSFs - see Fig.~\ref{fig:im_big}(b).}.
 
\begin{figure*}
\rotatebox{270}{\scalebox{0.4}{\includegraphics{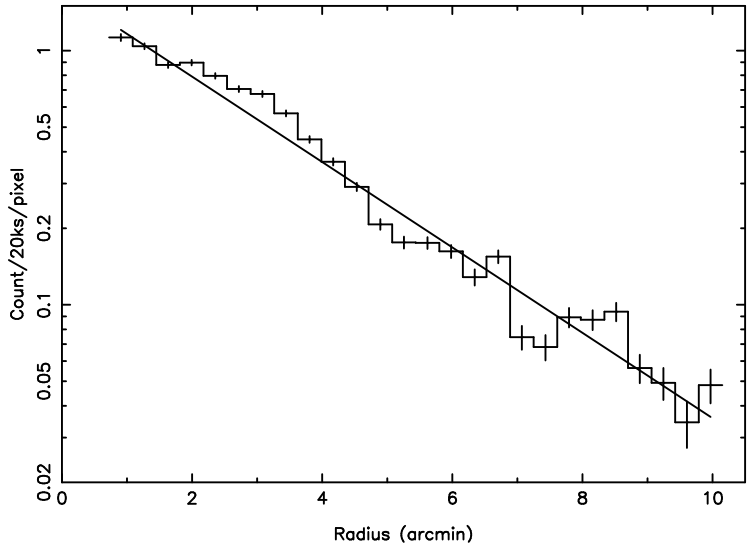}}}
\caption{The radial distribution of the soft-band (0.3-1 keV) pn+MOS count
rate. The sky background level has been subtracted.  The solid line shows an
exponential fall-off in the surface brightness with a scale length of 
2.6\arcm (5.4 kpc), as determined from optical B data (\citealt{okamura76}).}
\label{fig:radial}
\end{figure*}

The residual component is most prominent within the the central 5\arcm radius region,
although it can be traced out to 10\arcm in  the radial profile distribution
shown in Fig.~\ref{fig:radial}. There is a clear correlation with the spiral structure
of the galaxy, as is evident from a comparison with the {\it GALEX} near-UV 
(NUV; 2310 \AA) band image of M101 (\citealt{bianchi05}; \citealt{popescu05}) 
shown in Fig.~\ref{fig:im_big}(d).
 
\subsection{Spectral extraction of the diffuse component}
\label{sec:diff_extr}

The images in Fig.~\ref{fig:im_big} demonstrate the  existence of  an extensive  
residual component in the soft X-ray emission of M101. In order to investigate 
the spectral properties of this component we have extracted spectral data for 
the 5\arcm\ radius region, centred on the position of
the nucleus.  A total of nine individual diffuse emission  spectra were 
derived from the appropriate
event files (for the three EPIC instruments, over the three observations). In each 
case, regions around bright point sources were excluded using the mask described 
earlier (in practice this involved
the use of circular  exclusion cells of radius $12'' - 53''$, as an approximation
to the source mask used in the image analysis).  

Given the extended nature of the diffuse emission, the background subtraction
is more complicated than is usual with EPIC data.  The use of other \xmmn data 
sets deriving from ``blank-field'' observations is, in our view, not a good option 
due to the variation from field to field in the soft X-ray background from the sky 
- clearly an important consideration when attempting to measure a soft diffuse 
signal. Instead local background (LBG) spectra 
were extracted from each event file for the annular region from 
5\arcm to 10\arcm radius - see Fig.~\ref{fig:im_big}(a) - 
again with the regions around bright sources excluded.  We also extracted particle 
background (PBG) spectra from the out-of-field-of-view corner regions in each EPIC 
camera. The true background 
beneath the residual M101 emission in the central source region will be made up
primarily of (vignetted) photons from the X-ray sky 
and (non-vignetted) particles, which comprise the instrumental background. The 
LBG spectra will also have photon and particle components, but, as the LBG region 
is further 
off-axis than the source region, these components are not in the correct ratio 
for accurate background subtraction (since the signal reduction due to the vignetting 
is larger in the LBG region). However, a very good model of the true background 
can be made from a combination of the LBG and PBG spectra, provided the 
contributions of the two are appropriately scaled\footnote{The scale 
factors applied to the LBG and PBG spectra were respectively
$(A_{c}/A_{LBG})\times(V_{c}/V_{LBG})$ and $(A_{c}/A_{PBG})\times(1-V_{c}/V_{LBG})$. Here
$A_{c}$, $A_{LBG}$ and $A_{PBG}$ refer to the actual areas on the detector 
of the central, LBG and PBG extraction regions. Similarly $V_{c}$ and $V_{LBG}$
are the average vignetting factors within the central and LBG regions.}
to allow for the vignetting-corrected and actual areas of the central, LBG and 
PBG extraction regions.       
Using this approach, an appropriate composite background was determined for each
central region spectrum together with  the corresponding ARF and RMF files.
Finally for each dataset, the spectral channels were binned together to give a 
minimum of 20 counts per bin.

Given the radial profile in Fig.~\ref{fig:radial}, the above amounts to a differential 
measurement of the galaxy spectrum, which in effect compares the integrated signal
in the central 5\arcm region with the equivalent signal measured in the
5\arcm - 10\arcm annular region. 

\section{Results}
\label{sec:res}

\subsection{Luminosity and composition of the residual emission}
\label{sec:cont}

After the exclusion of the bright source regions and subtracting the mean sky surface 
brightness measured in the 5\arcm--10\arcmin annulus, the count rate in the 
central 5\arcm region is 0.28 pn+MOS $\rm ct~s^{-1}$ (0.3--1 keV). 
However, we need to apply three correction factors to obtain the ``true'' central 
5\arcm count rate. The first accounts for the loss of ``diffuse'' signal due to the 
source mask which blocks 12\% of the central area. The second factor allows for the 
differential nature of the measurement ({\it i.e.,} the fact that we extract the 
background from the 5\arcm-10\arcm annulus where the galaxy is still bright); on 
the basis of the radial profile determined earlier we estimate that this
effect suppresses the central count rate by $\approx 19\%$. Finally, the spillage 
of the bright source signal beyond the masked region (due to the 
extended PSF) contributes about 9\% of the measured residual count rate
(as determined from the source model image shown in Fig.~\ref{fig:im_big}(b)).   The 
correction of all three effects translates to a 28\% upward scaling of the residual 
count rate. If we adopt the best-fitting two-temperature model described 
in section \ref{sec:spec}, the count rate to flux conversion factor is 
1 pn+MOS $\rm ct~s^{-1}$ = $0.97 \times 10^{-12} \rm~erg~cm^{-2}~s^{-1}$ 
for the 0.5--2.0 keV band (fluxes
corrected for Galactic absorption).  The equivalent factors for the 0.3--1 keV and
0.45--1 keV bands are $1.16$ and $0.91 \times 10^{-12} 
\rm~erg~cm^{-2}~s^{-1}~(ct~s^{-1})^{-1}$ respectively.

The implied luminosity of the central emission of M101 after excluding 
bright sources is $2.1 \times 10^{39} \rm erg~s^{-1}$ in the 0.5--2 keV band.
However, we must take into account the fact that our source exclusion threshold is 
relatively high (roughly a factor 10 higher than that employed in the equivalent
\chandra study - \citealt*{kuntz03}). \chandra has shown that the log N - log S 
relation 
for relatively faint sources in the central region of M101 (after correction for 
background AGNs) is quite flat with a slope of $-0.80\pm 0.05$ in the integral
counts \citep{pence01}, implying that bright sources provide the dominant 
contribution to the discrete source luminosity of the galaxy. Using the count 
rate to flux conversion factor quoted earlier, our source exclusion threshold 
of 1.5 pn+MOS $\rm ct~ks^{-1}$ (0.3--1 keV) corresponds to a source luminosity 
of $9 \times 10^{36} \ergsec $ (0.5--2 keV),  although this could 
be an underestimate for the harder spectral forms more typical of bright X-ray 
binaries. In fact, a source exclusion threshold of $1 \times 10^{37} \ergsec $ 
appears to be a reasonable estimate based on the average count rate to flux conversion 
obtained when we  compare our measurements with those quoted in \citet{pence01} 
for the sub-set of common sources.  
As a check, at the latter luminosity threshold the \chandra source counts
predict $\approx 25$ sources within a central 5\arcm region, which is very consistent 
with the number of sources actually excluded in our study. Using the \chandra 
source counts, we find the integrated luminosity in discrete sources from 
our exclusion cut-off down to a factor 10 deeper (in effect the \chandra limit) 
amounts to about 20\% of the residual signal measured by {\it XMM-Newton}. 
In other words, our estimate 
of the ``diffuse'' luminosity of M101 (after allowing for the contribution of 
sources brighter than $\sim 10^{36} \ergsec $) is $1.7 \times 10^{39} \ergsec$ 
(0.5--2 keV) in good agreement with the value
reported by \citet{kuntz03} based on \chandra measurements
of the same region.

\subsection{Spectral properties of the residual emission}
\label{sec:spec}

\begin{figure*}
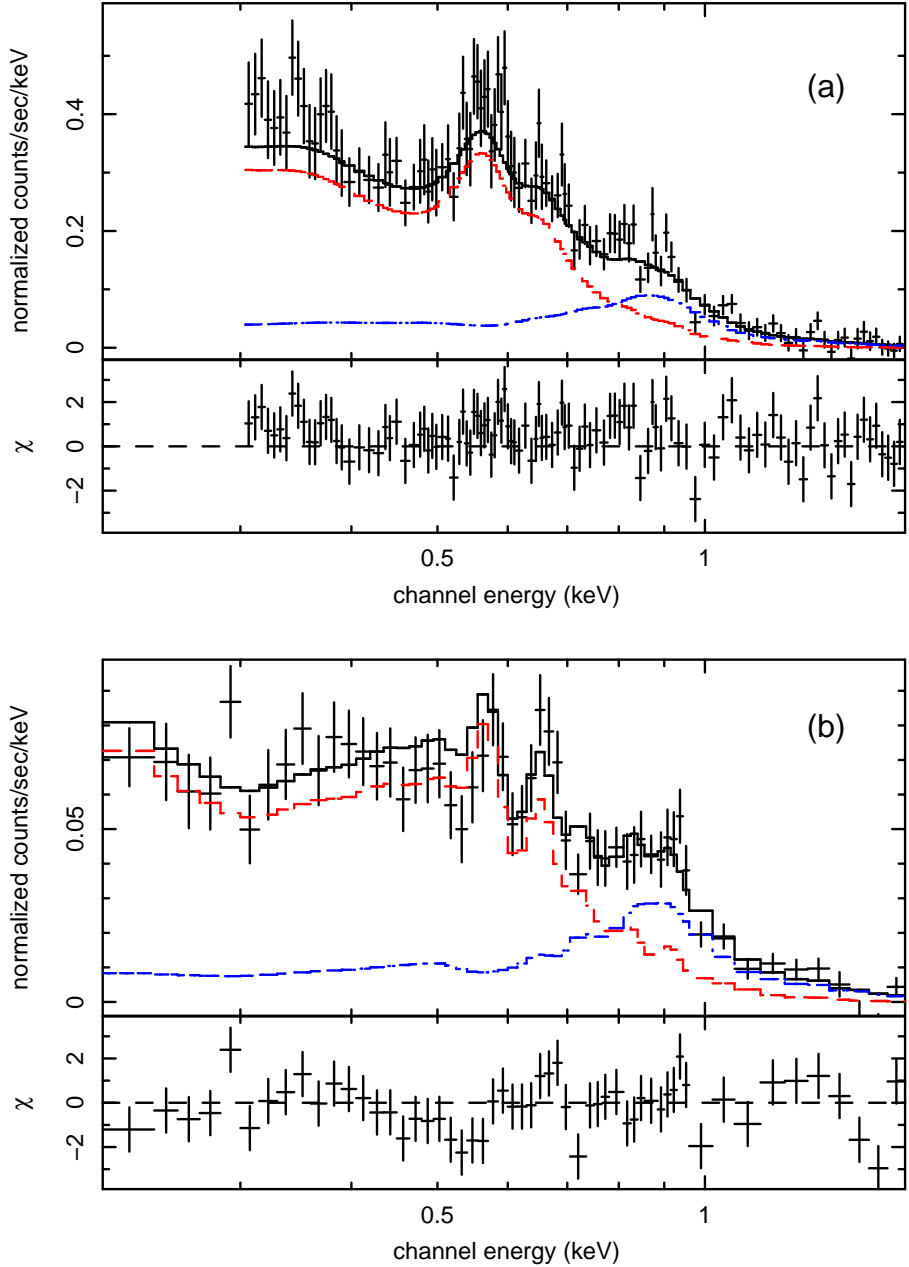

\centering
\rotatebox{270}{\scalebox{0.5}{\includegraphics{fig_3a.ps}}}
\rotatebox{270}{\scalebox{0.5}{\includegraphics{fig_3b.ps}}}
\caption{The EPIC spectra of the residual X-ray emission component in M101 
after the exclusion of bright sources. {\it (a)} The pn spectrum averaged
over the three observations.  {\it (b)} The combined MOS spectrum averaged
over the three observations.  In both cases the solid line corresponds to the 
best-fitting 2-component thermal spectrum with the dashed lines showing the 
individual contributions of the 0.2-keV and 0.7-keV components - see text.  
The $\chi^{2}$ residuals with respect to the best-fitting model are also 
shown. }

\label{fig:spec}
\end{figure*}

The nine epic spectra were fitted simultaneously using a variety of
spectral models, with identical model parameters applied to each
of the data sets. Visual inspection of the spectra
indicated almost no emission above $\approx 1.5$\,keV. No acceptable
fit could be obtained using just a single hot plasma component. 
However, a good fit (reduced $\chi^{2}$=0.87 for 2986 degrees of freedom) 
was obtained with a model incorporating two thermal (solar-abundance) Mekal 
components, with absorption attributable to the hydrogen column density of
our Galaxy in the direction of M101 (1.16$\times10^{20}$\,cm$^{-2}$; 
\citealt{dickey90}). 
The derived temperatures for the two components were 0.20$^{+0.01}_{-0.01}$ keV and 
0.68$^{+0.06}_{-0.04}$ keV (90\% confidence limits assuming 1 interesting parameter).
The data plus best fitting two-temperature
model are shown in Fig~\ref{fig:spec}. (In this figure we show the averages
of the three pn and the six MOS spectra for clarity of 
display; however, the spectral fitting results quoted here pertain to the fits
to the nine individual spectra).
The absorption-corrected flux in the 0.3--1 keV band derived from the spectral fits 
was 3.0 $\times10^{-13}$\,erg~cm$^{-2}$~s$^{-1}$ of which the hotter component 
contributes $18$\%. The corresponding values for the 0.5--2.0 keV 
band are 2.5 $\times10^{-13}$\,erg~cm$^{-2}$~s$^{-1}$ with the hotter component
contributing $27$\%.
When we allowed the metallicity of the two plasmas to vary from
unity ({\it i.e.,} solar values) we obtained a best fit of 0.044
(0.038-0.055) solar for the lower-temperature component and 0.17
(0.10-0.50) solar for the hotter component (with a reduced $\chi^{2}$ for the fit 
of 0.81). However, this determination of strongly sub-solar abundances 
is most probably an artifact of our attempt to fit relatively low spectral
resolution data pertaining to a complex multiphase plasma with a
simplistic spectral model, a problem which has in fact been well documented
in the literature ({\it e.g.,} \citealt{strickland00}; \citealt{wang01}). As noted by
\citet{kuntz03}, the abundances in the hot plasma in the central region of M101 
are more likely to be near to solar rather than grossly sub-solar.

We know that the measured \xmmn spectrum includes some residual  contamination
by the wings of the bright sources (estimated to be 9\% of the 
0.3--1 keV count rate) and also includes a contribution from the underlying 
unresolved source population (estimated to contribute $20\%$ of the 
0.5--2 keV flux). Since the sources may have somewhat harder spectra than
the diffuse emission, presumably they contribute preferentially to the 
higher temperature spectral component.  However, our derived temperatures 
of $\approx 0.2$ and $\approx 0.7$ keV
are in very good agreement with the results of \citet{kuntz03} based on
\chandra observations, suggesting that some fraction of the harder 
emission does have a diffuse origin. In the following analysis we assume
50\% of the measured hard emission is attributable to diffuse emission. 
We also scale the component normalisations obtained from the spectral 
fitting so as to correct for both the area of the source mask and 
the differential nature of the spectral measurement (as detailed in \S 3.1).

One can infer mean physical properties of the hot diffuse gas once
some assumptions have been made regarding the geometry of the diffuse
emission. Here we assume that the gas is contained within a cylindrical
region of radius of 10.5 kpc in the plane of the galaxy (matching a 
5\arcm angular extent) with a half-width perpendicular to the plane 
of 0.5 kpc (representing the extent of a putative shallow halo). 
Using this volume, the derived emission measure $\eta n^{2}_{e} V$ 
(where $\eta$ is the `filling factor' - the fraction of the total volume $V$ 
which is occupied by the emitting gas) can be used to infer the mean electron 
density $n_{e}$.  For the 0.2 keV plasma we find $n_{e} \approx 0.003$ 
$\eta^{-1/2}$\,cm$^{-3}$ compared to a value of $\approx 0.001$ 
$\eta^{-1/2}$\,cm$^{-3}$ for the 0.7 keV plasma, implying that these
two components are in rough pressure balance. The thermal energy 
residing in each of the plasma components is comparable and totals
$E_{th} \approx 6\times10^{55}$$\eta^{1/2}$\,erg. The cooling is dominated by the
line emission  with radiative cooling timescales of 
$t_{\mbox{\small soft}} \approx 1.8 \times 10^{8} \eta^{1/2}$\,yr 
and $t_{\mbox{\small hard}} \approx 1.5 \eta^{1/2}$\,Gyr for the
cooler and hotter components respectively.

\subsection{Morphology of the residual emission}
\label{sec:morph}

As previously noted, the spatial distribution of the residual soft X-ray emission in M101 
shows a good correlation with the optical/UV emission of the galaxy, at least 
over the central 5\arcm region. We have investigated this correlation using the
FUV (1530 \AA) and NUV (2310 \AA) images of M101 from the {\it GALEX} pipeline 
(\citealt{morrissey05}), together with images recorded by the \xmmn
Optical Monitor ({\it OM}) in the UVW1 ($\approx$ 2800 \AA), U, B and V filters
(\citealt{mason01}).
The latter were derived from data taken during observation~1 (UVW1) and observation 3 
(U,B,V) and subject to the standard SAS pipeline processing. 

To aid the comparison, the soft X-ray image of Fig.~\ref{fig:im_big}(c) was compressed 
in both coordinates by a factor 4 (to give $17.4\arcs \times 17.4\arcs$ pixels) 
and then lightly smoothed.  The UV and optical images were first rebinned into the same 
image format as the original soft X-ray data, blurred by a mask representative of 
the on-axis \xmmn soft X-ray PSF and then further compressed in an identical way to 
the soft X-ray data.

Fig.~\ref{fig:xuv} shows a detailed comparison of the 0.3--1 keV surface brightness 
measured in the central 5\arcm radius region of M101 with the corresponding
NUV, U and V band images.  A striking similarity between the soft X-ray
and U band images is immediately obvious. In particular, the ``S-feature'' which 
delineates the
inner spiral structure of M101 is very prominent in both cases.  
The fall-off in the surface brightness towards the edges of the 5\arcm field is also 
very similar in both wavebands, with the north-south asymmetry evident at the 
level of the lowest soft X-ray contour level also present in the U band image.
Interestingly, the correspondence becomes slightly less strong when the soft
X-ray image is compared to both the NUV and V band images. 
The NUV image is clearly totally dominated by bright HII regions 
and young star associations which together delineate the 
spiral arms of the galaxy, whereas in the soft X-ray image the general rise in 
surface brightness towards the nucleus has more relative impact. 
It would seem highly unlikely that these morphological trends could 
be induced by the increase in extinction as we move from the U band into
the NUV. In contrast, comparison with the V band data veers in the other
direction with the soft X-ray image showing more structure outside the central
2\arcm zone than the optical image.

\begin{figure*}
\centering
\caption{{\it (a)} The pn+MOS soft-band (0.3--1 keV) image of
the central 5\arcm radius region of M101. The scaling is linear.
The contour levels increase 
in steps of 0.3 pn+MOS $\rm ct~ks^{-1}~pixel^{-1}$ from a
starting level of 0.9 pn+MOS $\rm ct~ks^{-1}~pixel^{-1}$. 
{\it (b)} The {\it GALEX} NUV image of the same region
overlaid with the soft X-ray contours. The scaling is linear
in $\rm count~pixel^{-1}$ units.
{\it (c)} The  \xmmn {\it OM} U image of the same region
overlaid with the soft X-ray contours.The scaling is linear
in $\rm count~pixel^{-1}$ units.
{\it (d)} The  \xmmn {\it OM} V image of the same region
overlaid with the soft X-ray contours. The scaling is linear
in $\rm count~pixel^{-1}$ units.
{\tt Figures submitted to arXiv as jpgs.}
}
\label{fig:xuv}
\end{figure*}

\begin{figure*}
\centering
\rotatebox{270}{\scalebox{0.8}{\includegraphics{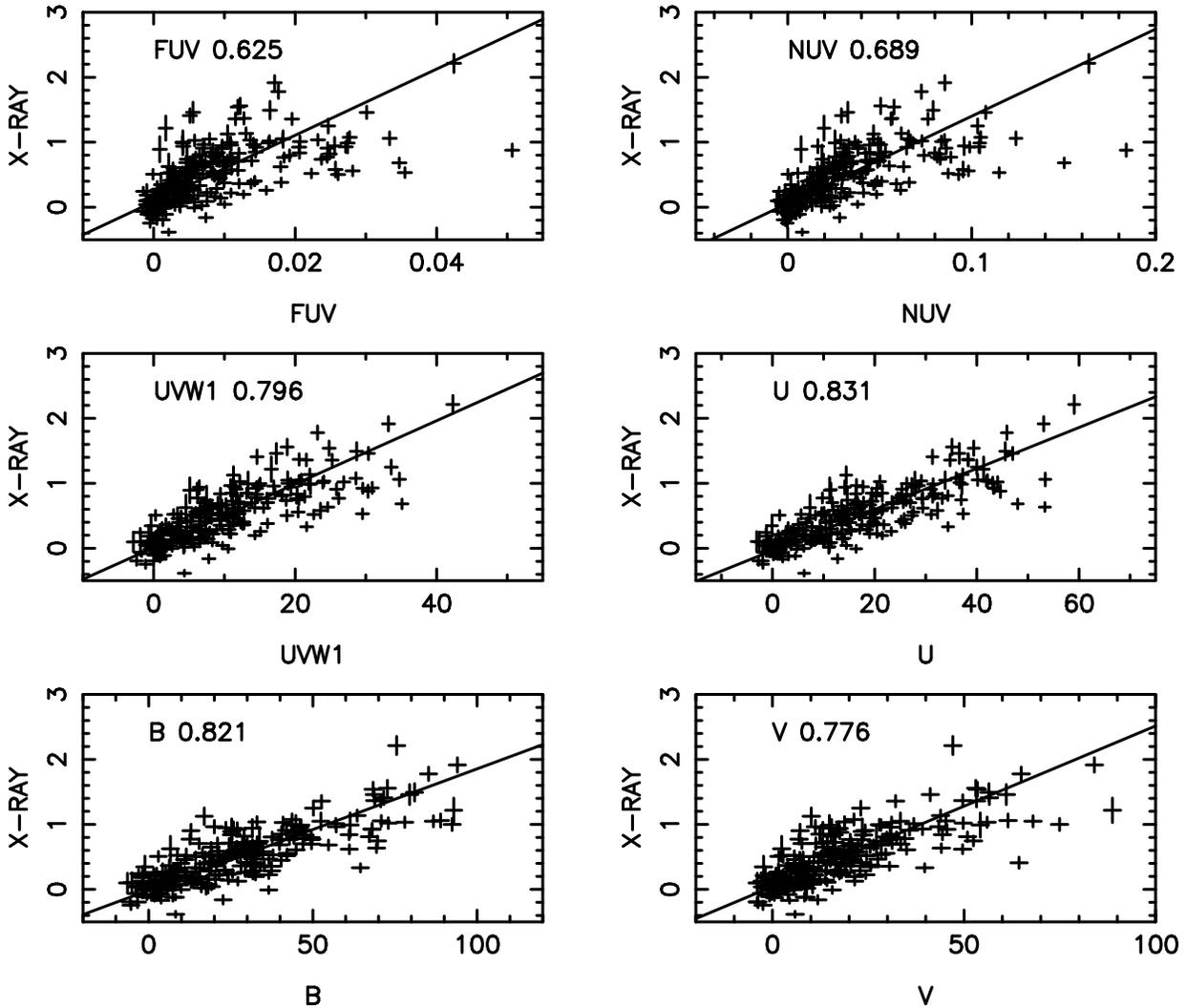}}}
\caption{The correlation of the surface brightness measured in the soft X-ray band
with that measured in other wavebands from the FUV through to V. The units in all 
cases are effectively counts per pixel. The numerical values quoted in each panel 
are the Pearson product moment correlation coefficient.
}
\label{fig:cor}
\end{figure*}

The above trends are further illustrated in Fig.~\ref{fig:cor} 
which shows the pixel-by-pixel correlation between the soft X-ray surface 
brightness and the optical/UV surface brightness in the various wavebands.
(here we use unsmoothed data with a pixel size of $34.8\arcs \times 34.8\arcs$, 
{\it i.e.} a factor of 8 compression of the original images).
The correlation, as measured by the Pearson product moment correlation 
coefficient, R, is at a maximum in the U band.

\section{Discussion}
\label{sec:disc}

After excluding bright point sources (with $L_X > 10^{37} \ergsec$)
we measure the residual soft X-ray luminosity of the inner (10.5 kpc radius) 
disk of M101 to be $L_X \approx 2.1 \times 10^{39} \ergsec$ (0.5--2 keV). 
Up to 20\% of this flux may originate in relatively bright point X-ray sources
(with  $L_X$ down to $10^{36} \ergsec$), which remain unresolved in
the \xmmn observations.  A further 5\% of the measured luminosity may 
be attributable to intermediate luminosity sources (with  $L_X$ down to
$10^{34}\ergsec$) and, at very faint levels, the integrated emission of 
dwarf stars may account for another 10\% contribution (\citealt{kuntz03}).
Nevertheless, it seems that the bulk of observed X-ray emission is likely 
to be diffuse in origin.

The EPIC spectrum of the residual emission can be modelled as the sum of
two thermal plasma components at temperatures of $\approx 0.2$ keV and 
$\approx 0.7$ keV with the harder component contributing $27\%$ of the flux in the 
0.5--2 keV band ($18\%$ in the 0.3--1 keV band). These spectral 
characteristics appear to be quite typical of the diffuse components seen
in normal and starburst galaxies ({\it e.g.,} \citealt{fraternali02};
\citealt{jenkins04b}; \citealt {jenkins05a}) and, 
albeit with some caveats, consistent 
with spectral measurements  for M101 reported in the literature 
(\citealt{snowden95}; \citealt{wang99};  \citealt{kuntz03}). 
 
Earlier studies have demonstrated that for normal spiral galaxies, the 
total soft X-ray luminosity correlates with the total infrared luminosity
(\citealt{read01}; \citealt{strickland04b}), the latter serving as a direct 
indicator of the current star formation rate. 
For the diffuse X-ray components, this is in line with the mechanism 
whereby the mechanical energy input from supernovae and the winds from 
young massive stars serves to heat the galactic ISM
to million-degree temperatures. In the case of M101 the required rate
of energy input is readily supplied by two supernovae per century 
(\citealt{matonick97}), if the efficiency of the conversion of the 
mechnical energy of the supernova explosion into X-rays 
is $10^{51}/E~\%$, where $E$ is the energy of the 
supernova in $\rm erg$.  

The  linkage of X-ray emission to recent star formation 
is further strengthened by the finding that in a number of galaxies
the X-ray morphology in the spiral arms matches that seen in the 
mid-infrared or H${\alpha}$ (\citealt{tyler04}).
A recent \chandra study has demonstrated that, in M101, the  
diffuse X-ray emission traces the spiral arms and is correlated with both the 
H${\alpha}$ and far-UV emission (\citealt{kuntz03}). In the same study 
there was no evidence for any significant trend of X-ray spectral hardness 
with radius\footnote{We have subsequently checked the \xmmn data for
spectral trends by extracting images in the 0.3--0.8 keV 
and  0.8--1.2 keV sub-bands.  The radial profiles of the residual X-ray emission 
in these sub-bands are consistent with the results obtained for the broader
0.3--1.0 keV band (see Fig.~\ref{fig:radial}) and within the errors, exhibit a 
constant ratio for radii in the range 1'-10'.  The \xmmn observations thus confirm 
that there is no significant radial dependence of spectral hardness in this Galaxy.} 
implying that the 0.2- and 0.7-keV spectral components are closely linked with 
one another and are both associated with star formation.

 In the present paper we 
have investigated this correlation in more detail by contrasting the X-ray 
morphology with that observed in far-UV through to V band.  
We find that the correlation with the X-ray morphology is quite strong for all 
bands considered (FUV, NUV, UVW1, U, B and V), although the best match is 
obtained with the U band data.  One interpretation of this result is that,
in broad terms, there are two underlying components present in the
X-ray surface brightness distribution, namely a smooth rise towards the
centre of the galaxy as reflected in the radial profile of the emission
(Fig.~\ref{fig:radial}) plus a superimposed X-ray spiral arm structure.
In moving from the FUV through to the V band, clearly the balance shifts between 
the extreme population I components (distributed in the spiral arms) and 
a more intermediate population (associated with the galactic disk) and
it is in the U band that this balance happens to best match the
X-ray morphology. 

In the disks of spiral galaxies star-formation is thought to be triggered 
by the passage of a spiral density wave through the ISM. 
The formation of massive young stars results in a large UV/far-UV flux
with reprocessing in the immediate environment subsequently giving rise to associated 
H${\alpha}$ and mid-IR emission. All of  these serve
as tracers of the spiral arms of the galaxy.  Population synthesis models 
predict that following a star-formation burst, the peak in the  UV/H${\alpha}$ 
production will last $\sim 3 \times 10^{6}$ yr, whilst the most massive stars 
complete their life cycle. The rate of energy deposition into the ISM
 from the subsequent supernovae rises at this time and remains fairly 
constant for $\sim 3 \times 10^{7}$ yr (\citealt{leitherer95}; \citealt{cervino02}). 
Given the lag between the peak in UV/H${\alpha}$ emission and the 
X-ray heating one might predict a spatial offset between the diffuse X-ray emission 
and the other spiral tracers, although as yet there is no clear observational
evidence for such an effect (\citealt{tyler04}). 

In the case of M101, we assume a flat rotation curve beyond a few kpc of the galaxy 
centre and a rotational velocity $v_{rot}\approx200~\rm~km~s^{-1}$ (\citealt{bosma81}). 
We further assume
that the spiral pattern corotates with the disk material at 3 radial scale lengths, 
{\it i.e.} at $r\approx15$ kpc. Then at $ r\approx7.5 $ kpc a 
delay of $3 \times 10^{7}$ yr 
translates to a rotational lag of $24^{\circ}$ or about 3 kpc in a direction perpendicular
to the spiral arm.  The effective pixel size in our multi-waveband correlation  is $35\arcs$ 
or 1.2 kpc, implying that a delay in the X-ray production following the passage of a 
spiral density wave might just be measurable in the current data. Unfortunately the
restricted  signal to noise and limited clarity of the spiral arm features
in the X-ray image mitigates against the detection of such an effect.
Earlier we estimated a radiative cooling timescale for the more prominent 0.2-keV 
plasma component to be $t_{\mbox{\small soft}} \approx 1.8 \times 10^{8}$\,yr
for a filling factor $\eta \sim 1$. Clearly this timescale is inconsistent 
with the presence of reasonably narrow spiral features in the soft X-ray image.
A very small filling factor, {\it e.g.} $\eta \sim 0.01$, is required to match
the properties of X-ray spiral arms in M101, suggesting a very clumpy thin-disk 
distribution. In fact a reasonable scenario is that the spiral arms in X-rays 
are delineated by a combination of truly diffuse emission, possibly in the form
of hot gas bubbles and superbubbles, plus contributions from individual SNRs and 
concentrations of unresolved discrete sources. 
As noted by \citet{kuntz03}, the available
constraints suggest that superbubbles in the disk of M101 may  have similar 
properties to the Galactic Loop I superbubble (\citealt{egger95}),
although they are not individually resolvable even in \chandra data.
Where there is no confinement by chimneys or similar structures in the ISM, 
energy losses arising from adiabatic expansion of the hot gas in the disk
into the lower halo of M101 may also help localise the spiral arm component
({\it nb.} in M101 the sound speed  of the soft component is comparable to the escape 
velocity from the disk - \citealt{kuntz03}).

The second spatial component of the X-ray emission considered above broadly 
follows the distribution of the optical light attributable to the intermediate 
disk stellar population.  The integrated X-ray emission of dwarf stars will 
presumably follow a disk distribution but this is unlikely to represent a 
substantial contribution (10\% overall in the 0.5--2 keV band but, given the 
typical $kT \sim 0.8$ keV temperature of dwarf-star coronae,
falling to $\approx 5\%$ in the softer 0.3--1 keV band; \citealt{kuntz03}).
Other contributions to this smoother distribution might be expected from supernovae 
in the interarm regions associated with disk population stars and X-ray emitting gas 
which has accumulated in a shallow halo in M101 as a result of galactic chimney/fountain
activity. In the case of the latter, gas which has cooled substantially due to adiabatic 
expansion may, as a result of a frozen-in non-ionization equilibrium,
exhibit an emission spectrum dominated by oxygen lines, which mimics that expected 
from a $0.2$ keV plasma (\citealt{breitschwerdt99}). The fact that the X-ray 
surface brightness increases towards the centre of M101 is most naturally
explained in terms of an increase in activity per unit disk area,  with the filling 
factor of the extended z-height component possibly approaching unity near the 
centre of the galaxy. With the latter assumption, the implied electron density and 
pressure in the inner halo of M101 is comparable to that inferred for the centre of 
Galactic Loop I and the Local Hot Bubble.

\section{Conclusions}
\label{sec:conc}

This is the third and final paper in a series presenting the results of \xmmn observations
of the nearby face-on Scd  spiral galaxy M101. Here we focus on the spatial and spectral
properties of the  galaxy, when bright X-ray sources with $L_X > 10^{37} \ergsec$ are 
removed using an appropriate point source mask. 

The residual soft X-ray luminosity of the central (10.5 kpc radius) region of M101 was
measured as $L_X \approx 2.1 \times 10^{39} \ergsec$ (0.5--2 keV), the bulk of which 
appears to originate as diffuse emission. We find a two-temperature model best fits
the spectral data with the derived temperatures of 0.20$\pm$0.01 keV and 
0.68$^{+0.06}_{-0.04}$ keV, typical of the diffuse components seen
in other normal and starburst galaxies. 

In line with earlier studies, we find that the observed X-ray surface brightness 
distribution is well correlated with images recorded in optical/UV wavebands.
In particular the detection of spiral arms in X-rays establishes a close link between 
the X-ray emission and recent starformation. Closer investigation suggests that the 
X-ray morphology may comprise both a spiral arm component and a smoother disk component. 
In spiral galaxies, star formation is thought to be triggered by the passage of a 
spiral density wave through the ISM. In principle one might observe a lag between more 
immediate star-formation indicators such as the UV/far-UV flux or H${\alpha}$ and 
the heating of the interstellar medium to X-ray temperatures as result of 
subsequent supernova. The \xmmn data show no evidence for such an effect
but it may be observable in future high resolution and high signal-to-noise data.
Whereas the spiral arm component, on the basis of the radiative cooling
timescale of the 0.2-keV plasma, may be deduced to have a clumpy, low-z distribution,
the smoother disk component may represent longer-lived  X-ray gas with a relatively
large filling factor, which has been transported to the lower halo of M101 via the
galactic chimney/fountain mechanism.

Future progress in understanding the X-ray properties of normal spiral
galaxies will no doubt follow from very intensive studies of individual
galaxies ({\it e.g.,} the recent 1 Ms observation of M101 with \chandra - Kuntz et al.
in preparation) and by applying appropriately optimised analysis procedures to 
samples of nearby galaxies, as is now possible using archival datasets.

\section*{Acknowledgments}

LPJ, TPR and RAO acknowledge PPARC support at various junctures during 
this project. We should also like to thank the anonymous referee for
some helpful comments and suggestions.

\label{lastpage}

{}


\section*{APPENDIX: The Point Source Catalogue}

Within the \d25 circle of M101, 108 sources were detected in observation~1, 91 
in observation~2 and the same number in  observation~3. Ninety  sources were 
detected  in at  
least two  of the observations. For  completeness, Table~\ref{table:srclist} lists  
the additional 21 sources detected in Observation~2 and another 9 sources  
detected in Observation~3. The format of the table is identical to
that of Table 1 of Paper II and the numbering  scheme is a continuation
of  the original list  of 108 sources. 

The most  notable new  source to  appear in  the second  and third
observations  is  source \#122,  which  corresponds  to  the  supersoft  transient  P98
\citep{mukai03}. The results from the \xmmn and \chandra observations
of this object are reported in \citet{kong04} and \citet{mukai05}.

\begin{table*}
\caption{Additional sources detected within the \d25 circle of M101 in 
observations 2 (\#109--129) and 3 (\#130--138).}
 \centering
{\scriptsize
  \begin{tabular}{@{}lcccccccccccc@{}}
\hline

Src & XMMU         & r$_{1\sigma}$ & \multicolumn{3}{c}{pn count rate (count ks$^{-1}$)} & \multicolumn{3}{c}{MOS count rate (count ks$^{-1}$)} & F$_X$   & L$_X$      & HR1            & HR2   \\
    &                  & ($^{\prime\prime}$)   & S      & M            & H            & S            & M            & H            &    &                   &          &     \\
(1) & (2)              & (3)    & \multicolumn{3}{c}{(4)}              & \multicolumn{3}{c}{(5)}                    & (6)            & (7)            & (8)            & (9)  \\

\hline

109  &	J140213.8+542158  &  1.95  &  {\bf 3.1$\pm$0.9}  &  0.1$\pm$0.3  &  1.7$\pm$0.8  &  {\bf 0.8$\pm$0.2}  &  {\bf 0.9$\pm$0.2}  &  0.7$\pm$0.2  &  1.58$\pm$0.27  &  0.98$\pm$0.17  &  -0.47$\pm$0.12  &  0.21$\pm$0.16  \\
110  &	J140237.7+542733  &  2.98  &  -            &  -            &  -            &  0.4$\pm$0.2  &  {\bf 0.9$\pm$0.2}  &  0.3$\pm$0.2  &  0.90$\pm$0.30  &  0.56$\pm$0.19  &  0.41$\pm$0.24   &  -0.52$\pm$0.28  \\
111  &	J140238.0+541215  &  2.31  &  -            &  -            &  -            &  0.5$\pm$0.2  &  {\bf 1.0$\pm$0.3}  &  0.3$\pm$0.2  &  1.11$\pm$0.32  &  0.69$\pm$0.20  &  0.34$\pm$0.21   &  -0.49$\pm$0.26  \\
112  &	J140246.7+541454  &  2.01  &  0.0$\pm$0.3  &  {\bf 2.8$\pm$0.6}  &  1.3$\pm$0.5  &  0.0$\pm$0.1  &  {\bf 0.7$\pm$0.2}  &  0.4$\pm$0.2  &  1.00$\pm$0.21  &  0.62$\pm$0.13  &  1.00$\pm$0.14   &  -0.32$\pm$0.15  \\
113  &	J140251.4+540901  &  2.93  &  -            &  -            &  -            &  {\bf 1.2$\pm$0.3}  &  {\bf 1.2$\pm$0.3}  &  0.4$\pm$0.2  &  1.56$\pm$0.39  &  0.97$\pm$0.24  &  0.00$\pm$0.18   &  -0.49$\pm$0.26  \\
114  &	J140257.9+542655  &  2.07  &  1.0$\pm$0.4  &  1.4$\pm$0.5  &  1.5$\pm$0.5  &  0.2$\pm$0.1  &  0.5$\pm$0.2  &  {\bf 0.8$\pm$0.2}  &  1.27$\pm$0.22  &  0.79$\pm$0.13  &  0.25$\pm$0.20   &  0.16$\pm$0.15  \\
115  &	J140302.0+541336  &  2.00  &  1.8$\pm$0.5  &  0.8$\pm$0.4  &  0.0$\pm$0.2  &  {\bf 0.6$\pm$0.2}  &  0.2$\pm$0.1  &  0.1$\pm$0.1  &  0.40$\pm$0.12  &  0.25$\pm$0.08  &  -0.43$\pm$0.17  &  -0.67$\pm$0.41  \\
116  &	J140304.0+541040  &  2.07  &  2.8$\pm$0.8  &  1.2$\pm$0.5  &  0.0$\pm$0.4  &  0.5$\pm$0.2  &  {\bf 0.8$\pm$0.2}  &  0.1$\pm$0.2  &  0.64$\pm$0.18  &  0.39$\pm$0.11  &  -0.07$\pm$0.15  &  -0.76$\pm$0.28  \\
117  &	J140306.9+541007  &  2.28  &  -            &  -            &  -            &  0.3$\pm$0.2  &  0.0$\pm$0.1  &  {\bf 1.2$\pm$0.3}  &  1.79$\pm$0.47  &  1.11$\pm$0.29  &  -1.00$\pm$0.89  &  1.00$\pm$0.22  \\
118  &	J140308.3+543229  &  2.01  &  -            &  -            &  -            &  {\bf 1.4$\pm$0.3}  &  0.0$\pm$0.1  &  0.0$\pm$0.0  &  0.60$\pm$0.17  &  0.38$\pm$0.10  &  -1.00$\pm$0.13  &  0.00$\pm$0.00  \\
119  &	J140320.0+542034  &  2.93  &  -            &  -            &  -            &  {\bf 0.7$\pm$0.2}  &  0.1$\pm$0.1  &  0.0$\pm$0.1  &  0.35$\pm$0.14  &  0.22$\pm$0.09  &  -0.70$\pm$0.24  &  -1.00$\pm$1.36  \\
120  &	J140324.0+542336  &  2.58  &  {\bf 2.5$\pm$0.5}  &  0.4$\pm$0.3  &  0.0$\pm$0.1  &  -            &  -            &  -            &  0.34$\pm$0.11  &  0.21$\pm$0.07  &  -0.75$\pm$0.17  &  -0.83$\pm$0.69  \\
121  &	J140327.5+541909  &  1.82  &  {\bf 2.1$\pm$0.5}  &  1.1$\pm$0.3  &  0.8$\pm$0.3  &  0.4$\pm$0.1  &  {\bf 0.5$\pm$0.1}  &  0.4$\pm$0.1  &  0.88$\pm$0.14  &  0.55$\pm$0.09  &  -0.14$\pm$0.14  &  -0.12$\pm$0.16  \\
122  &	J140332.4+542103  &  1.73  &  {\bf 5.5$\pm$0.7}  &  0.1$\pm$0.2  &  0.6$\pm$0.3  &  {\bf 0.9$\pm$0.2}  &  0.0$\pm$0.0  &  0.0$\pm$0.1  &  0.55$\pm$0.10  &  0.34$\pm$0.06  &  -0.98$\pm$0.04  &  0.64$\pm$0.45  \\
123  &	J140334.3+540930  &  1.95  &  {\bf 4.2$\pm$0.9}  &  {\bf 2.7$\pm$0.7}  &  1.9$\pm$0.8  &  {\bf 1.0$\pm$0.3}  &  0.5$\pm$0.2  &  0.9$\pm$0.3  &  1.91$\pm$0.32  &  1.19$\pm$0.20  &  -0.26$\pm$0.13  &  0.04$\pm$0.17  \\
124  &	J140354.1+541108  &  2.20  &  -            &  -            &  -            &  {\bf 1.4$\pm$0.3}  &  0.0$\pm$0.1  &  0.1$\pm$0.2  &  0.79$\pm$0.27  &  0.49$\pm$0.17  &  -1.00$\pm$0.17  &  1.00$\pm$1.74  \\
125  &	J140355.8+542100  &  1.68  &  {\bf 6.3$\pm$0.8}  &  {\bf 2.2$\pm$0.5}  &  0.7$\pm$0.4  &  {\bf 0.9$\pm$0.2}  &  {\bf 0.7$\pm$0.2}  &  0.1$\pm$0.1  &  1.03$\pm$0.14  &  0.64$\pm$0.09  &  -0.38$\pm$0.08  &  -0.62$\pm$0.16  \\
126  &	J140357.2+541011  &  2.11  &  {\bf 5.5$\pm$1.1}  &  2.3$\pm$0.7  &  1.1$\pm$0.7  &  0.6$\pm$0.3  &  0.4$\pm$0.3  &  1.0$\pm$0.3  &  1.67$\pm$0.32  &  1.03$\pm$0.20  &  -0.38$\pm$0.14  &  0.02$\pm$0.21  \\
127  &	J140401.2+542344  &  2.59  &  {\bf 2.3$\pm$0.5}  &  0.0$\pm$0.2  &  0.3$\pm$0.4  &  -            &  -            &  -            &  0.44$\pm$0.23  &  0.27$\pm$0.14  &  -1.00$\pm$0.16  &  1.00$\pm$1.09  \\
128  &	J140429.8+542235  &  2.19  &  -            &  -            &  -            &  0.3$\pm$0.2  &  {\bf 0.8$\pm$0.2}  &  0.6$\pm$0.2  &  1.25$\pm$0.30  &  0.78$\pm$0.19  &  0.43$\pm$0.23   &  -0.18$\pm$0.22  \\
129  &	J140435.4+542101  &  1.59  &  {\bf 12.9$\pm$1.3} &  0.0$\pm$0.2  &  0.0$\pm$0.2  &  -            &  -            &  -            &  1.41$\pm$0.20  &  0.88$\pm$0.12  &  -1.00$\pm$0.03  &  -1.00$\pm$1.00  \\

\hline

130  &  J140229.9+542243  &  1.54  &  {\bf 14.0$\pm$1.4}  &  0.0$\pm$0.3  &  0.0$\pm$0.4  &  -            &  -            &  -            &  1.44$\pm$0.26  &  0.89$\pm$0.16  &  -1.00$\pm$0.04  &   0.00$\pm$0.00  \\
131  &  J140230.0+541611  &  2.05  &  3.6$\pm$1.1   &  {\bf 4.4$\pm$1.0}  &  3.2$\pm$1.1  &  {\bf 1.2$\pm$0.3}  &  1.2$\pm$0.4  &  0.7$\pm$0.4  &  2.30$\pm$0.42  &  1.43$\pm$0.26  &   0.07$\pm$0.14  &  -0.20$\pm$0.17  \\
132  &  J140243.6+542007  &  2.38  &  {\bf 2.1$\pm$0.6}   &  0.4$\pm$0.3  &  0.4$\pm$0.4  &  -            &  -            &  -            &  0.48$\pm$0.24  &  0.30$\pm$0.15  &  -0.70$\pm$0.25  &   0.04$\pm$0.71  \\
133  &  J140250.7+542857  &  3.01  &  {\bf 2.5$\pm$0.7}   &  0.0$\pm$0.1  &  1.3$\pm$0.7  &  -            &  -            &  -            &  1.00$\pm$0.37  &  0.62$\pm$0.23  &  -1.00$\pm$0.09  &   1.00$\pm$0.17  \\
134  &  J140300.6+540958  &  2.24  &  {\bf 4.3$\pm$1.0}   &  2.0$\pm$0.7  &  1.0$\pm$0.8  &  -            &  -            &  -            &  1.29$\pm$0.45  &  0.80$\pm$0.28  &  -0.36$\pm$0.18  &  -0.36$\pm$0.38  \\
135  &  J140312.2+541754  &  2.19  &  {\bf 3.8$\pm$0.7}   &  0.3$\pm$0.3  &  0.1$\pm$0.2  &  -            &  -            &  -            &  0.50$\pm$0.14  &  0.31$\pm$0.09  &  -0.84$\pm$0.13  &  -0.45$\pm$0.76  \\
136  &  J140314.4+542132  &  2.79  &  -             &  -            &  -            &  {\bf 0.6$\pm$0.2}  &  0.2$\pm$0.1  &  0.2$\pm$0.1  &  0.57$\pm$0.19  &  0.35$\pm$0.12  &  -0.49$\pm$0.25  &  -0.15$\pm$0.48  \\
137  &  J140355.8+540856  &  2.10  &  {\bf 4.0$\pm$0.9}   &  {\bf 2.9$\pm$0.8}  &  1.3$\pm$0.7  &  -            &  -            &  -            &  1.57$\pm$0.44  &  0.97$\pm$0.27  &  -0.17$\pm$0.17  &  -0.39$\pm$0.27  \\
138  &  J140415.6+540948  &  1.75  &  {\bf 7.2$\pm$1.1}   &  {\bf 4.5$\pm$0.9}  &  1.5$\pm$0.9  &  {\bf 2.0$\pm$0.4}  &  {\bf 1.0$\pm$0.3}  &  0.6$\pm$0.3  &  2.12$\pm$0.31  &  1.31$\pm$0.19  &  -0.26$\pm$0.09  &  -0.41$\pm$0.17  \\

\hline
\end{tabular}
}
\begin{tabular}{@{}l@{}}
(1) source number; (2) XMMU source designation (J2000 coordinates); (3) 1$\sigma$ error radius (including a 1.5 arcsecond systematic error);\\
(4 \& 5) source count rates in soft (0.3--1\,keV), medium (1--2\,keV) \& hard (2--6\,keV) bands for the pn and MOS cameras, with the\\ significant source detections ($>4\sigma$) highlighted in bold; (6) source flux in units of $10^{-14} \ergcms$ in the broad (0.3--6\,keV) band; \\(7) source luminosity in units of $10^{38} \ergsec$ in the 0.3--6\,keV band (assuming a distance to M101 of 7.2\,Mpc); (8 \& 9) soft (HR1) and \\hard (HR2) hardness ratios.\\
\end{tabular}
\label{table:srclist}
\end{table*}

\end{document}